\documentclass[twocolumn,preprintnumbers,noshowpacs,noshowkeys]{revtex4}
\usepackage{amssymb}
\usepackage{graphicx}
\usepackage{dcolumn}
\usepackage{bm}
\usepackage{amsmath}
\usepackage{color}

\begin{document}

\title{Controllable Fano resonance and fast to slow light in a hybrid
semiconductor/superconductor ring device mediated by Majorana fermions}
\author{Hua-Jun Chen}
\email{chenphysics@126.com}
\affiliation{School of Mechanics and Photoelectric Physics, Anhui University of Science
and Technology, Huainan Anhui 232001, China}

\begin{abstract}
We demonstrate theoretically the Fano resonance and the conversion from fast
to slow light in a hybrid quantum dot-semiconductor/superconductor ring
device, where the QD is coupled to a pair of MFs appearing in the hybrid S/S
ring device. The absorption spectra of the weak probe field can exhibit a
series of asymmetric Fano line shapes and their related propagation
properties such as fast and slow light effects are investigated based on the
hybrid system for suitable parametric regimes. The positions of the Fano
resonances can be determined by the parameters, such as different detuning
regimes and QD-MFs coupling strengths. Further, the transparency windows
(the absorption dip approaches zero) in the probe absorption spectra are
accompanied by the rapid dispersion, which indicates the slow or fast light
effect, and tunable fast-to-slow light propagation (or vice versa) can be
achieved by controlling different parameter regimes. Our study may provide
an all-optical means to investigate MFs and open up promising applications
in quantum information processing based on MFs in solid state devices.
\end{abstract}

\maketitle




\section{INTRODUCTION}

\ The coherent interplay of laser field with multiple-level atom systems can
induce electromagnetically induced transparency (EIT), where an opaque
medium can be made transparent due to quantum interference between two
quantum pathways in $\Lambda $-type atoms \cite{FleischhauerM} leading to a
symmetric transparency window. The EIT plays an important role in modern
quantum optics, and the EIT technique has been applied extensively to
control the group velocity of light \cite{HauLV,BudkerD}, realize the
storage of quantum information \cite{LiuC}, obtain the enhanced nonlinear
processes \cite{HarrisSE}, and achieve optical switch \cite{ShenJQ1} and
quantum interference \cite{ShenJQ2}. Different from EIT, which presents a
symmetric transparency window, the Fano line profile shows an asymmetry
shape caused by the scattering of light amplitude when the condition of
observing EIT is not meet and an extra frequency detuning is introduced.
Fano resonance was first demonstrated by Fano due to the scattering
interference of a narrow discrete resonance with a broad spectral line or
continuum \cite{FanoU,MiroshnichenkoAE}, which plays a key role in the
fields of graphene systems \cite{GuoJ}, photonics crystal systems \cite%
{NojimaS}, and plasmonics \cite{ArtarA}, and also induces many potential
applications in sensors \cite{LeeKL}, enhanced light emission \cite{ZhouZK},
and slow light \cite{WuC}. Due to the multiple EIT phenomenon is the
manifestation of Fano resonances, and then Fano resonances can be elucidated
from the output field with the multiple EIT approach.

On the other hand, the quickly developing experimental detection schemes of
Majorana fermions (MFs) have witnessed great progress in the past two
decades in condensed matter systems \cite{AliceaJ}, including hybrid
semiconducting nanowire (atomic chains)/superconductor structure \cite%
{MourikV,DasA,DengMT,Nadj-PergeS2,ChenJ}, iron-based superconductor \cite%
{YinJX}, topological structure \cite{AlbrechtSM,SunHH}, and quantum
anomalous Hall insulator--superconductor structure \cite{HeQL}. MFs are
exotic particles whose antiparticle are themself $\gamma =\gamma ^{\dagger }%
\mathbf{\ }$obeying non-Abelian statistics, which then may reach subsequent
potential applications in topological quantum computation and quantum
information processing \cite{ElliottSR}. Up to now, several typical means
for detecting MFs have also been presented experimentally, such as zero-bias
peaks (ZBPs) in tunneling spectroscopy \cite%
{MourikV,DasA,DengMT,Nadj-PergeS2,ChenJ}, fractional a.c. Josephson effect%
\cite{Rokhinson}, Coulomb blockade spectroscopy experiment \cite{AlbrechtSM}%
, and spin-polarized scanning tunneling microscopy \cite{JeonS}. Further,
benefit from significant progress in modern nanoscience and nanotechnology,
quantum dots (QDs) \cite{Urbaszek}, provide the fantastic medium to detect
MFs both theoretically \cite{LiuDE,Flensberg1,Leijnse,Pientka,Sau1} and
experimentally \cite{DengMT2}.

Rather than previous\ schemes for probing MFs in electrical domain, we once
have proposed an all-optical means to detect MFs with QD considered as a
two-level system (TLS) and driven by two-tone fields \cite%
{Chen01,Chen02,Chen03}, which may afford a potential supplement to detect
MFs. However, there is no research work focusing on Fano resonance mediated
by MFs in a hybrid semiconductor/superconductor (S/S) ring device \cite%
{Chen03}. Additionally, achieving the switch of fast-to-slow light based on
Fano resonance has never attracted enough academic attention. In the present
paper, we first demonstrate that the probe absorption spectra of the QD can
display the switch from EIT to Fano resonances induced by MFs under
different detuning regimes in the hybrid S/S ring device, which can be
explained by the interference effect in terms of the dressed states. The
Fano resonances can be effectively tuned and the probe absorption spectra
can display a series of asymmetric Fano line shapes under different
parameter regimes including QD-MFs coupling strengths $\beta _{1}$ and $%
\beta _{2}$, the exciton-pump field detuning $\Delta _{c}$, and the
Majorana-pump field detuning $\Delta _{M}$. Secondly, we investigate the
slow light effect by numerically calculating the group delay of the probe
field around the transparency window accompanied by the steep phase
dispersion, and we find a tunable and controllable fast-to-slow light
propagation (and vice versa) can be achieved with manipulating the parameter
regimes.

\section{Physical Pattern and Dynamical Equation}

\begin{figure}[tbp]
\includegraphics[width=8.8cm]{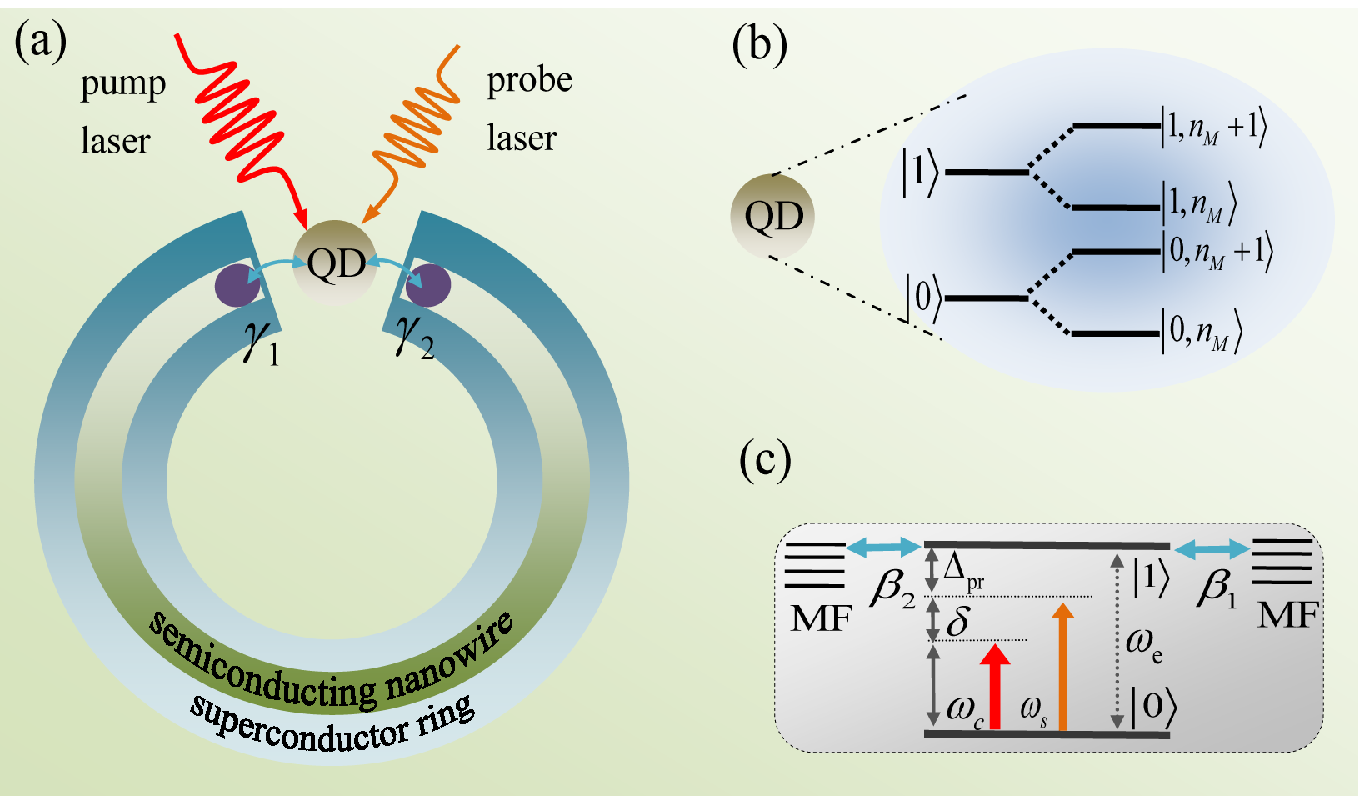}
\caption{(a) draws the schematic setup of hybrid QD-S/S ring device, where a
QD driven by two-tone fields coupled to a couple of MFs appearing in\ the
end of semiconducting ring. (b) The QD is considered as a TLS and coupled to
a coupled of MFs inducing four coupled states $\left\vert
0,n_{M}\right\rangle $, $\left\vert 0,n_{M}+1\right\rangle $, $\left\vert
1,n_{M}\right\rangle $ and $\left\vert 1,n_{M}+1\right\rangle $, where $%
n_{M} $ indicates the number states of the MFs. (c) The energy-level diagram
of the QD coupled to a couple of MFs.}
\end{figure}

Figure 1 gives the schematic setup that will be studied in this paper, in
which a QD coupled to a pair of MFs appearing in the hybrid S/S ring device,
and the Hamiltonian is given by \cite%
{LiuDE,Flensberg1,Leijnse,Pientka,Sau1,Chen03,Boyd}%
\begin{eqnarray}
H &=&\hbar \omega _{e}S^{z}+i\epsilon _{M}\gamma _{1}\gamma _{2}/2+i\hbar
\beta _{1}(S^{-}-S^{\dag })\gamma _{1}  \notag \\
&&+i\hbar \beta _{2}(S^{-}-S^{\dag })\gamma _{2}-\mu E_{c}(S^{\dag
}e^{-i\omega _{c}t}+S^{-}e^{i\omega _{c}t})  \notag \\
&&-\mu E_{s}(S^{\dag }e^{-i\omega _{s}t}+S^{-}e^{i\omega _{s}t}),
\end{eqnarray}%
where the first term indicates the Hamiltonian of the QD with the exciton
frequency $\omega _{e}$. To distinguish from previous theoretical schemes
for detecting MFs, where QD is consider as a single resonant level with
spin-singlet state \cite{LiuDE,Flensberg1,Leijnse,Pientka,Sau1}, here we
consider the QD is a TLS including the ground state $\left\vert
0\right\rangle $ and the single exciton state $\left\vert 1\right\rangle $
\cite{ZrennerA,StuflerS} at low temperature. We introduce the pseudospin
operators $S^{z}$ and $S^{\pm }$ with the commutation relations $\left[
S^{z},S^{\pm }\right] =\pm S^{\pm }$ and $\left[ S^{+},S^{-}\right] =2S^{z}$
to describe the TLS QD.

The second term gives the interaction of a pair of MFs emerging in the
hybrid S/S ring device. We use Majorana operators $\gamma _{1}$ and $\gamma
_{2}$ to describe MFs satisfying the relation $\gamma ^{\dagger }=\gamma $
and $\gamma ^{2}=1$ as they are their own antiparticle. Here, $\epsilon
_{M}=\hbar \omega _{M}\sim e^{-l/\xi }$ is the coupling energy with $l$
being the semiconductor nanowire length and $\xi $ the superconducting
coherent length with Majorana frequency $\omega _{M}$. Obviously, when the
semiconductor nanowire length $l$ is large enough, the coupling energy $%
\epsilon _{M}$ approach zero. Therefore, it is necessary to discuss the two
conditions, i.e., the coupled Majorana edge states $\epsilon _{M}\neq 0$ and
the uncoupled Majorana edge states $\epsilon _{M}=0$.

The third and the fourth terms describe a pair MFs $\gamma _{1}$ and $\gamma
_{2}$ coupled to the QD with the coupling strengths $\beta _{1}$ and $\beta
_{2}$, and the coupling strengths relate to the distance of the QD and S/S
ring device. For simplicity, we introduce the regular fermion creation
(annihilation) operator $f^{\dag }$($f$), and then Majorana operator $\gamma
$ can be transformed into the regular fermion operator $f$ with the relation
of $\gamma _{1}=f^{\dag }+f$ and $\gamma _{2}$ $=i(f^{\dag }-f)$. Therefore,
the third and the fourth terms in Eq. (1) can be rewritten as $i\hbar \beta
_{1}(S^{-}f^{\dag }-S^{+}f)-\hbar \beta _{2}(S^{-}f^{\dag }+S^{+}f)$ in the
rotating wave approximation \cite{RidolfoA}, where the non-conservation
terms of energy $i\hbar \beta _{1}(S^{-}f-S^{+}f^{+})$ and $\hbar \beta
_{2}(S^{-}f+S^{+}f^{+})$ are neglected due to their effect are too small to
be considered in our theoretical treatment.

The last two terms in Eq. (1) indicate the interactions between the QD and
two laser fields including a strong pump field with frequency $\omega _{c}$
and a weak probe field with frequency $\omega _{s}$ simultaneously
irradiating to the QD, where $\mu $ is the electric dipole moment of the
exciton, $E_{c}$ and $E_{s}$ are the slowly varying envelope of the pump
field and probe field, respectively.

In a frame rotating with the frequency $\omega _{c}$ of the pump field, the
Hamiltonian of the system in Eq. (1) can be rewritten as%
\begin{eqnarray}
H &=&\hbar \Delta _{c}S^{z}+\hbar \Delta _{M}(f^{\dag }f-1/2)+i\hbar \beta
_{1}(S^{-}f^{\dag }-S^{+}f)  \notag \\
&&-\hbar \beta _{2}(S^{-}f^{\dag }+S^{+}f)-\hbar \Omega _{c}(S^{+}+S^{-})
\notag \\
&&-\mu E_{s}(S^{+}e^{-i\delta t}+S^{-}e^{i\delta t}),
\end{eqnarray}%
where $\Delta _{c}=\omega _{e}-\omega _{c}$ is the exciton-pump field
detuning, $\Delta _{M}=\omega _{M}-\omega _{c}$ is the Majorana-pump field
detuning, and $\delta =\omega _{s}-\omega _{c}$ is the probe-pump detuning. $%
\Omega _{c}=\mu E_{c}/\hbar $ is the Rabi frequency of the pump field. The
Heisenberg-Langevin equations of the operators for the resonators, including
the corresponding noise and damping terms, can be written as follows \cite%
{Chen03,Walls}:
\begin{eqnarray}
\partial _{t}S^{z} &=&-\Gamma _{1}(S^{z}+1/2)-\beta _{1}(S^{-}f^{\dag
}+S^{\dag }f)  \notag \\
&&-i\beta _{2}(S^{-}f^{\dag }+S^{\dag }f)+i\Omega _{c}(S^{\dag }-S^{-})
\notag \\
&&+\frac{i\mu E_{s}}{\hbar }(S^{\dag }e^{-i\delta t}-S^{-}e^{i\delta t}),
\end{eqnarray}%
\begin{eqnarray}
\partial _{t}S^{-} &=&-(i\Delta _{c}+\Gamma _{2})S^{-}+2(\beta _{1}-i\beta
_{2})S^{z}f  \notag \\
&&-2i\Omega _{c}S^{z}-\frac{2i\mu E_{s}S^{z}e^{-i\delta t}}{\hbar }+\tau
_{in}\text{,}
\end{eqnarray}%
\begin{equation}
\partial _{t}f=-(i\Delta _{M}+\kappa _{M}/2)f+(\beta _{1}+i\beta
_{2})S^{-}+\xi \text{,}
\end{equation}%
where $\Gamma _{1}$ ($\Gamma _{2}$) is the exciton relaxation rate (exciton
dephasing rate), and $\kappa _{M}$ is the decay rate of the MFs. $\tau _{in}$
is the $\delta $-correlated Langevin noise operator with zero mean, and $\xi
$ is Langevin force arising from the interaction between the Majorana modes
and the environment.

Using $S^{z}=S_{0}^{z}+\delta S^{z}$, $S=S_{0}+\delta S$ and $f=f_{0}+\delta
f$, Eqs. (3)-(5) can be divided into the steady parts and the fluctuation
ones. Substituting the division forms into Eqs. (3)-(5) and setting all the
time derivations at the steady parts to be zero, we obtain the steady state
solutions of the variables determining the steady-state population inversion
($w_{0}=2S_{0}^{z}$) of the exciton, which obeys the following equation%
\begin{gather}
\Gamma _{1}(w_{0}+1)[w_{0}^{2}(\beta _{1}^{2}+\beta
_{2}^{2})^{2}-w_{0}(\beta _{1}^{2}+\beta _{2}^{2})(\Gamma _{2}\kappa
_{M}-2\Delta _{c}\Delta _{M})  \notag \\
+(\Delta _{c}^{2}+\Gamma _{2}^{2})(\Delta _{M}^{2}+\kappa
_{M}^{2}/4)]+4\Omega _{c}^{2}w_{0}\Gamma _{2}(\Delta _{M}^{2}+\kappa
_{M}^{2}/4)=0.
\end{gather}%
As all the pump fields are assumed to be sufficiently strong, all the
operators can be identified with their expectation values under the
mean-field approximation $\left\langle Qc\right\rangle =\left\langle
Q\right\rangle \left\langle c\right\rangle $ \cite{AgarwalGS}. After being
linearized by neglecting nonlinear terms in the fluctuations, the Langevin
equations for the expectation values%
\begin{eqnarray}
\left\langle \partial _{t}\delta S^{z}\right\rangle &=&-\Gamma
_{1}\left\langle \delta S^{z}\right\rangle -(\beta _{1}+i\beta
_{2})(S_{0}\left\langle \delta f^{\dag }\right\rangle +f_{0}^{\ast
}\left\langle \delta S^{-}\right\rangle )  \notag \\
&&-(\beta _{1}-i\beta _{2})(S_{0}^{\ast }\left\langle \delta f\right\rangle
+f_{0}\left\langle \delta S^{\dag }\right\rangle )  \notag \\
&&+i\Omega _{pu}(\left\langle \delta S^{\dag }\right\rangle -\left\langle
\delta S^{-}\right\rangle )  \notag \\
&&+\dfrac{i\mu E_{s}}{\hbar }[S_{0}^{\ast }e^{-i\delta t}-S_{0}e^{i\delta t}]%
\text{,}
\end{eqnarray}%
\begin{eqnarray}
\left\langle \partial _{t}\delta S^{-}\right\rangle &=&-(i\Delta
_{pu}+\Gamma _{2})\left\langle \delta S^{-}\right\rangle +w_{0}(\beta
_{1}-i\beta _{2})\left\langle \delta f\right\rangle  \notag \\
&&+2[f_{0}(\beta _{1}-i\beta _{2})-i\Omega _{c}]\left\langle \delta
S^{z}\right\rangle  \notag \\
&&-\frac{i\mu w_{0}E_{s}}{\hbar }e^{-i\delta t}\text{,}
\end{eqnarray}%
\begin{equation}
\left\langle \partial _{t}\delta f\right\rangle =-(i\Delta _{M}+\kappa
_{M}/2)\left\langle \delta f\right\rangle +(\beta _{1}+i\beta
_{2})\left\langle \delta S^{-}\right\rangle \text{,}
\end{equation}%
which is a set of nonlinear equations and the steady-state response in the
frequency domain is composed of many frequency components. To solve these
equations, we make the ansatz \cite{Boyd} as $\left\langle \delta
O\right\rangle =O_{+}e^{-i\delta t}+O_{-}e^{i\delta t}$ ($O=S^{z},S^{-},f$),
and substitute them into Eqs. (7)-(9) with ignoring the second-order terms
and working to the lowest order in $E_{s}$ but to all orders in $E_{c}$, we
obtain the linear optical susceptibility as $\chi _{eff}^{(1)}(\omega
_{s})=\mu S_{+}(\omega _{s})/E_{s}=(\mu ^{2}/\hbar \Gamma _{2})\chi
^{(1)}(\omega _{s})$, and $\chi ^{(1)}(\omega _{s})$ is given by%
\begin{equation}
\chi ^{(1)}(\omega _{s})=\frac{[{\epsilon }_{7}\Pi _{1}(\Lambda
_{4}+\epsilon _{3}\Pi _{2})-iw_{0}\Lambda _{4}]\Gamma _{2}}{{\Lambda
_{1}\Lambda _{4}+}\Pi _{1}\Pi _{2}\epsilon _{3}\epsilon _{4}},
\end{equation}%
where $\Pi _{1}=2[(\beta _{1}-i\beta _{2})f_{0}-i\Omega _{c}]$, $\Pi
_{2}=2[(\beta _{1}+i\beta _{2})f_{0}+i\Omega _{c}]$, ${\epsilon }_{1}=(\beta
_{1}+i\beta _{2})/[i(\Delta _{M}-\delta )+\kappa _{M}/2]$, ${\epsilon }%
_{2}=(\beta _{1}+i\beta _{2})/[i(\Delta _{M}+\delta )+\kappa _{M}/2]$, ${%
\epsilon }_{3}=[i\Omega _{c}-(\beta _{1}-i\beta _{2})f_{0}-(\beta
_{1}+i\beta _{2})S_{0}{\epsilon }_{2}^{\ast }]/(\Gamma _{1}-i\delta )$, ${%
\epsilon }_{4}=[i\Omega _{c}+(\beta _{1}+i\beta _{2})f_{0}^{\ast }+(\beta
_{1}-i\beta _{2})S_{0}^{\ast }{\epsilon }_{1}]/(\Gamma _{1}-i\delta )$, ${%
\epsilon }_{5}=[i\Omega _{c}-(\beta _{1}-i\beta _{2})f_{0}-(\beta
_{1}+i\beta _{2})S_{0}{\epsilon }_{1}^{\ast }]/(\Gamma _{1}+i\delta )$, ${%
\epsilon }_{6}=[i\Omega _{c}+(\beta _{1}+i\beta _{2})f_{0}^{\ast }+(\beta
_{1}-i\beta _{2})S_{0}^{\ast }{\epsilon }_{2}]/(\Gamma _{1}+i\delta )$, ${%
\epsilon }_{7}=iS_{0}^{\ast }/(\Gamma _{1}-i\delta )$, ${\epsilon }%
_{8}=iS_{0}/(\Gamma _{1}+i\delta )$, ${\Lambda _{1}=i(\Delta _{c}-\delta )}%
+\Gamma _{2}-w_{0}(\beta _{1}-i\beta _{2}){\epsilon }_{1}+\Pi _{1}\epsilon
_{4}$, ${\Lambda _{2}=-i(\Delta _{c}-\delta )}+\Gamma _{2}-w_{0}(\beta
_{1}+i\beta _{2}){\epsilon }_{1}^{\ast }-\Pi _{2}\epsilon _{5}$, ${\Lambda
_{3}=i(\Delta _{c}+\delta )}+\Gamma _{2}-w_{0}(\beta _{1}-i\beta _{2}){%
\epsilon }_{2}+\Pi _{1}\epsilon _{6}$, ${\Lambda _{4}=-i(\Delta _{c}+\delta )%
}+\Gamma _{2}-w_{0}(\beta _{1}+i\beta _{2}){\epsilon }_{1}+\Pi _{2}\epsilon
_{3}$. The imaginary and real parts of $\chi ^{(1)}(\omega _{s})$ indicate
absorption and dispersion, respectively.

Based on the hybrid QD-S/S ring device, we can determine the light group
velocity as \cite{Harris,Bennink} $v_{g}=c/[n+\omega _{s}(dn/d\omega _{s})]$
where $n\approx 1+2\pi \chi _{eff}^{(1)}$. Therefore%
\begin{eqnarray}
c/v_{g} &=&1+2\pi Re\chi _{eff}^{(1)}(\omega _{s})_{\omega _{s}=\omega _{e}}
\notag \\
&&+2\pi \omega _{s}Re(d\chi _{eff}^{(1)}/d\omega _{s})_{\omega _{s}=\omega
_{e}}\text{.}
\end{eqnarray}%
Obviously, when $Re\chi _{eff}^{(1)}(\omega _{s})_{\omega _{s}=\omega }=0$,
the dispersion is steeply positive or negative, and the group velocity is
significantly reduced or increased. Thus we define the group velocity index $%
n_{g}$ as%
\begin{eqnarray}
n_{g} &=&\dfrac{c}{v_{g}}-1=\dfrac{c-v_{g}}{v_{g}}  \notag \\
&=&\frac{2\pi \omega _{e}\rho \mu ^{2}}{\hbar \Gamma _{2}}Re(\frac{d\chi
_{eff}^{(1)}}{d\omega _{s}})_{\omega _{s}=\omega _{e}}  \notag \\
&=&\Gamma _{2}\Sigma Re(\frac{d\chi _{eff}^{(1)}}{d\omega _{s}})_{\omega
_{s}=\omega _{e}}\text{,}
\end{eqnarray}%
where $\Pi =2\pi \omega _{e}\rho \mu ^{2}/\hbar \Gamma _{2}$. One can
observe the slow light when $n_{g}>0$, and the superluminal light when $%
n_{g}<0$ \cite{Boyd01}.

The parameter values used in the paper \cite%
{MourikV,DasA,DengMT,Nadj-PergeS2,Chen03,Wilson-RaeI}: the QD-MFs coupling
strength $\beta _{1}=\beta _{2}=0.05$ GHz, the decay rate of the MFs $\kappa
_{M}=0.1$ MHz, $\Gamma _{1}=0.3$ GHz, $\Gamma _{2}=0.15$ GHz, and $\Omega
_{pu}^{2}=0.005($GHz$)^{2}$.

\section{MMIT, Fano resonance, and tunable switch of fast-to-slow light}

\subsection{Case A: the exciton-pump field detuning $\Delta _{c}=0$ and $%
\Delta _{c}=0.5$ GHz under uncoupled Majorana modes $\Delta _{M}=0$ ($%
\protect\epsilon _{M}=0$), respectively.}

In Fig. 1, we first consider the condition of uncoupled Majorana modes,
i.e., $\Delta _{M}=0$ or $\epsilon _{M}=0$ under the exciton-pump field
detuning $\Delta _{c}=0$. Then the Hamiltonian of the coupled QD-MFs system
reduce to $H_{MFs-QD}=i\hbar \beta _{1}(S^{-}f^{\dag }-S^{+}f)-\hbar \beta
_{2}(S^{-}f^{\dag }+S^{+}f)$ which is similar to J-C Hamiltonian of standard
model, and no the term of $\hbar \Delta _{M}(f^{\dag }f-1/2)$ in Eq.(1).
Therefore, the probe absorption spectra will present symmetric splitting.
Figure 2(a) gives the imaginary part of the dimensionless susceptibility Im$%
\chi ^{(1)}$ which indicates the probe absorption spectra of the probe field
as a function of the probe detuning $\Delta _{s}=\omega _{s}-\omega _{e}$
under four different QD-MFs coupling strengths $\beta _{2}$ under $\beta
_{1}=0.05$ GHz. The black curve in Fig.2(a) shows the result of only
considering the QD-MFs coupling $\beta _{1}$, and the probe absorption
spectrum shows a symmetric splitting. With increasing the QD-MFs coupling
strengths $\beta _{2}$, the distance of the splitting is enhanced
significantly, and the bigger QD-MFs coupling $\beta _{2}$ induced larger
peak width of splitting. Figure 2(b) shows the dispersion of the probe light
in the case $\Delta _{c}=0$ under uncoupled Majorana modes for several
different QD-MFs coupling strengths $\beta _{2}$ with $\beta _{1}=0.05$ GHz.
When increasing the coupling strength $\beta _{2}$ from $\beta _{2}=0$ to $%
\beta _{2}=3\beta _{1}$, the dispersion of the probe light varies from
negative to positive around $\Delta _{s}=0$, which results in positive group
delay or slow light propagation through the system. Therefore, the change of
dispersion from negative to positive slope with manipulating the coupling
strength $\beta _{2}$ corresponds to the control of fast-to-slow light
propagation. In Fig.2(c), we further present three different QD-MFs coupling
strength including ($\beta _{1}=0.05$ GHz, $\beta _{2}=0.05$ GHz), ($\beta
_{1}=0.07$ GHz, $\beta _{2}=0.03$ GHz), and ($\beta _{1}=0.09$ GHz, $\beta
_{2}=0.01$ GHz), which also manifests a symmetric splitting in the probe
absorption spectrum, and Fig.2(d) gives the dispersion of the probe light in
this condition. The physical origin of such results are due to the coherent
interaction between the QD and MFs, and here we introduce the dressed state
theory between the QD and MFs\ to interpret this physical phenomena. As the
QD is consider as a TLS with the ground state $\left\vert 0\right\rangle $
and exciton state $\left\vert e\right\rangle $, when QD couple to MFs, the
two-level QD will modify by the number states of the MFs $n_{M}$ ($n_{M}$ is
the number states of the MFs) generating the Majorana dressed states $%
\left\vert 0,n_{M}\right\rangle $, $\left\vert 0,n_{M}+1\right\rangle $, $%
\left\vert 1,n_{M}\right\rangle $, $\left\vert 1,n_{M}+1\right\rangle $. The
left sharp peak of splitting in the probe absorption spectra indicate the
transition from $\left\vert 0\right\rangle $ to $\left\vert
1,n_{M}\right\rangle $, and the right sharp peak is due to the transition of
$\left\vert 0\right\rangle $ to $\left\vert 1,n_{M}+1\right\rangle $. In
addition, the probe absorption spectra show the analogous phenomenon of
electromagnetically induced transparency (EIT) (the symmetric splitting)
\cite{FleischhauerM} in $\Lambda $-type atoms systems. Due to it is majorana
modes induced such phenomenon, and then we term the phenomenon as Majorana
modes induced transparency (MMIT). Obviously, the absorption dip approaches
zero around $\Delta _{s}=0$ under uncoupled Majorana modes $\epsilon _{M}=0$%
, which means the input probe field is transmitted to the coupled system
without experiencing any absorption.
\begin{figure}[tbp]
\includegraphics[width=9cm]{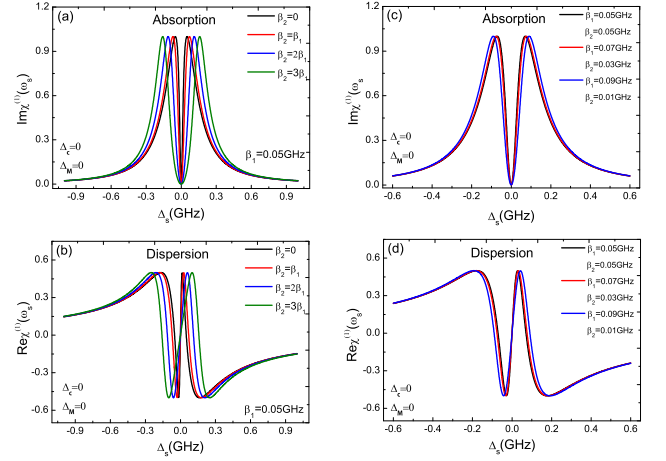}
\caption{(a) The probe absorption spectra of the probe field as a function
of the probe detuning $\Delta _{s}=\protect\omega _{s}-\protect\omega _{e}$
under four different QD-MFs coupling strengths $\protect\beta _{2}$ under $%
\protect\beta _{1}=0.05$ GHz and (b) gives their dispersion of the probe
light in this condition. (c) plots three different QD-MFs coupling strengths
$\protect\beta _{1}$ and $\protect\beta _{2}$ and (d) is the related
dispersion. The detuning conditions are $\Delta _{c}=0$ and $\Delta _{M}=0$.}
\end{figure}

When switching the exciton-pump field detuning $\Delta _{c}$ from $\Delta
_{c}=0$ and $\Delta _{c}=0.5$ GHz, the probe absorption spectra experience
the symmetric splitting (i.e., EIT) to unsymmetric splitting (i.e., Fano
resonance) under the condition of uncoupled majorana modes $\epsilon _{M}=0$
as shown in Fig. 3. Figure 3(a) displays the probe absorption spectra
presenting Fano resonance under several different QD-MFs coupling strengths $%
\beta _{2}$ with $\beta _{1}=0.05$ GHz, and Fig. 3(b) is its detail around $%
\Delta _{s}=0$. It is obvious that the unsymmetric splitting is increased
with bigger QD-MFs coupling strengths $\beta _{2}$. Figure 3(c) plots the
dispersion of the probe light for four different QD-MFs coupling strengths $%
\beta _{2}$ and Fig. 3(d) is its detail. We see that there is a steep
positive slope around $\Delta _{s}=0$, which signifies the potential of
ultraslow light achievement. Compared with Fig.2(a) and Fig.2(c), the
absorption dip in Fig. 3(a) also approaches zero around $\Delta _{s}=0$,
which is also analogous EIT without experiencing any absorption. However,
the physical regimes are different from in Fig. 2. The results are ascribed
to the destructive quantum interference effect between Majorana modes and
the beat frequency $\delta $ of the two input laser fields irradiating on
the QD. Once the beat frequency $\delta $ approximates to the resonance
frequency $\omega _{M}$ of Majorana modes, Majorana modes begin oscillating
coherently resulting in Stokes-like ($\Delta _{S}=\omega _{c}-\omega _{M}$)
and anti-Stokes-like ($\Delta _{A-S}=\omega _{c}+\omega _{M}$) scattering of
light from the QD. When the condition is highly off-resonant ($\Delta
_{c}=0.5$ GHz), the Stokes-like scattering is strongly suppressed, and only
leaving the anti-Stokes-like field interferes with the near-resonant probe
field modifying the probe absorption spectra. As a result, the probe
absorption spectra present zero absorption, i.e., MMIT.
\begin{figure}[tbp]
\includegraphics[width=9cm]{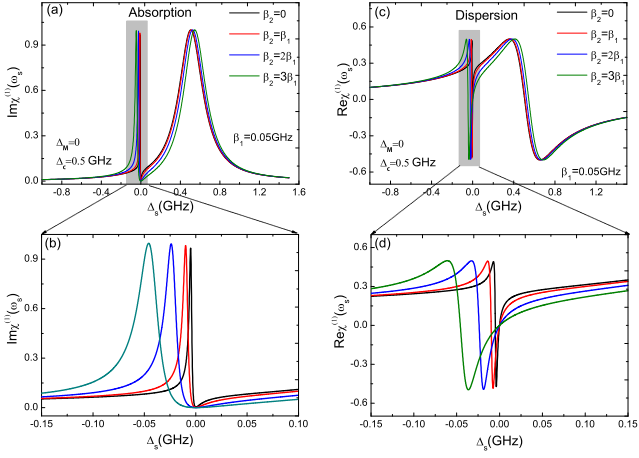}
\caption{(a) The probe absorption spectra under several different QD-MFs
coupling strengths $\protect\beta _{2}$ with $\protect\beta _{1}=0.05$ GHz,
and (b) is its detail aroud $\Delta _{s}=0$. (c) The dispersion of the probe
light for four different QD-MFs coupling strengths $\protect\beta _{2}$, and
(d) is its detail. The detuning conditions are $\Delta _{c}=0.5$ GHz and $%
\Delta _{M}=0$.}
\end{figure}

No matter what regimes induce the zero absorption in the probe absorption
spectra, once MMIT appear in the hybrid coupled system, the remarkable
phenomena of slow light or fast light can appear in the hybrid system. The
dispersion of the probe light in Fig.2(b) and Fig.3(d) show the dispersion
experiencing negative to positive slope with controlling the coupling
strength $\beta _{2}$ corresponds to the control of fast-to-slow light
propagation. In Fig.4(a), we consider the condition of $\Delta _{c}=0$ and $%
\Delta _{M}=0$. Figure 4(a) plots the group velocity index $n_{g}$ of probe
laser (in the unit of $\Pi $) as a function of one of QD-MFs coupling
strengths $\beta _{1}$ under two cases of $\beta _{2}=0$ and $\beta _{2}=0.1$
GHz, respectively. It is obvious that group velocity index $n_{g}$
experiences the change of positive-negative-positive, which corresponds to
the slow-fast-slow light in the both two conditions, respectively. In
Fig.4(b), we consider the condition of $\Delta _{c}=0.5$ GHz and $\Delta
_{M}=0$. We further show the group velocity index $n_{g}$ as a function of $%
\beta _{1}$ under the cases of $\beta _{2}=0$ and $\beta _{2}=0.1$ GHz,
respectively, which presents $n_{g}$ experiencing the switch from fast to
slow light.
\begin{figure}[tbp]
\includegraphics[width=9cm]{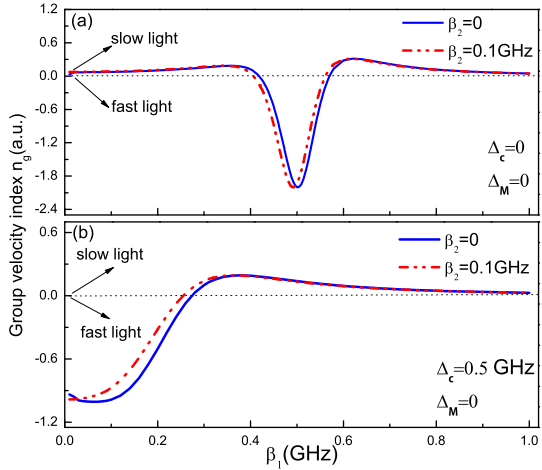}
\caption{(a) The group velocity index $n_{g}$ of probe laser as a function
of $\protect\beta _{1}$ under two cases of $\protect\beta _{2}=0$ and $%
\protect\beta _{2}=0.1$ GHz under $\Delta _{c}=0$ and $\Delta _{M}=0$,
respectively. (b) The group velocity index $n_{g}$ as a function of $\protect%
\beta _{1}$ under the cases of $\protect\beta _{2}=0$ and $\protect\beta %
_{2}=0.1$ GHz under $\Delta _{c}=0.5$ GHz and $\Delta _{M}=0$, respectively.}
\end{figure}

\subsection{Case B: the exciton-pump field detuning $\Delta _{c}=0$ under
coupled Majorana modes $\protect\epsilon _{M}\neq 0$}

When the semiconductor nanowire length $l$ is large enough, the coupling
energy $\epsilon _{M}$ approach zero. For the realistic device, length $l$
is limited, and then $\epsilon _{M}\neq 0$. Therefore, it is necessary to
discuss the condition, of the coupled Majorana edge states $\epsilon
_{M}\neq 0$. In Fig.5, we discuss the Fano resonance on the condition of
coupled Majorana modes ($\epsilon _{M}\neq 0$). Figure 5(a) plots the probe
absorption spectra under $\Delta _{c}=0$ for several different QD-MFs
coupling strengths $\beta _{2}$ with $\beta _{1}=0.05$ GHz, and Fig. 5(b) is
its detail. With increasing the coupling strengths $\beta _{2}$, the
splitting of the two peaks is broadened. In this case, the transparency
windows locate at $\Delta _{s}=-0.5$ GHz, which is different from in
Fig.3(a) locating at $\Delta _{s}=0$. Figure 5(c) plots the dispersion of
the probe light with four coupling strengths $\beta _{2}$ and Fig. 5(d) is
its detail, which has the same process of evolution as in Fig.3(c).
\begin{figure}[tbp]
\includegraphics[width=9cm]{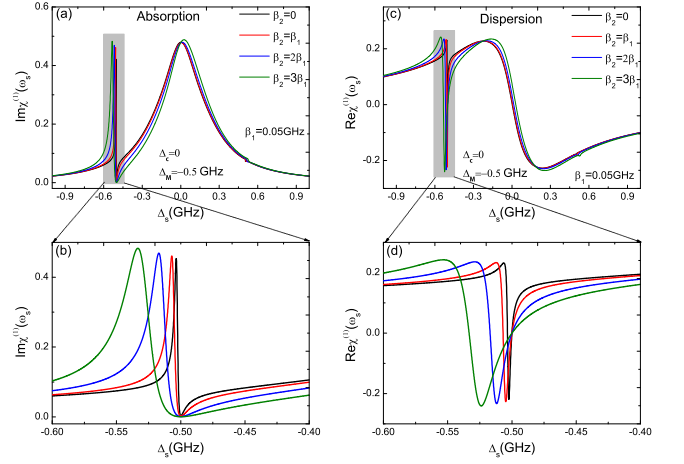}
\caption{(a) The probe absorption spectra under four QD-MFs coupling
strengths $\protect\beta _{2}$ with $\protect\beta _{1}=0.05$ GHz, and (b)
is its detail aroud $\Delta _{s}=0$. (c) The dispersion of the probe light
for four different QD-MFs coupling strengths $\protect\beta _{2}$, and (d)
is its detail. The detuning conditions are $\Delta _{c}=0$ and $\Delta
_{M}=-0.5$ GHz.}
\end{figure}

Figure 6 plots the group velocity index $n_{g}$ of probe laser under
different parameter regimes. In Fig.6(a), we investigate the group velocity
index $n_{g}$ as a function of $\beta _{1}$ under the condition of $\beta
_{2}=0$ and $\beta _{2}=0.1$ GHz, respectively, which presents the same
evolution as in Fig.4(a) indicating the slow-fast-slow light in the hybrid
system. We further discuss several different QD-MFs coupling strengths, such
as ($\beta _{2}=0.05$ GHz, $\beta _{2}=0$), ($\beta _{2}=0.05$ GHz, $\beta
_{2}=0.05$ GHz), ($\beta _{2}=0.07$ GHz, $\beta _{2}=0.03$ GHz), and ($\beta
_{2}=0.09$ GHz, $\beta _{2}=0.01$ GHz), which influences the group velocity
index $n_{g}$. Figure 6(b) plots the group velocity index $n_{g}$ of probe
laser versus the Rabi frequency $\Omega _{pu}^{2}$ of the pump field in the
case of $\beta _{1}=0.05$ GHz and $\beta _{2}=0$ under the condition of $%
\Delta _{c}=0$ and $\epsilon _{M}\neq 0$. One can see from Fig. 6(b) that
the group velocity index $n_{g}$ is positive when the Rabi frequency varies,
which represent the slow light effect. Figure 6(c) presents the group
velocity index $n_{g}$ \textit{vs.} Rabi frequency $\Omega _{pu}^{2}$ under
other three conditions, i.e., ($\beta _{2}=0.05$ GHz, $\beta _{2}=0.05$
GHz), ($\beta _{2}=0.07$ GHz, $\beta _{2}=0.03$ GHz), and ($\beta _{2}=0.09$
GHz, $\beta _{2}=0.01$ GHz). When the QD is close to one MF (such as $\gamma
_{1}$, and then $\beta _{1}>\beta _{2}$), the experience of the group
velocity index $n_{g}$ is different from the case of $\beta _{1}=\beta _{2}$%
, which can reach the switch of slow to fast light. The QD-MFs coupling
strength can be controlled via adjusting the distance of the QD and the
hybrid S/SR device. Therefore, with controlling the QD-MFs coupling
strengths, the fast-to-slow light or vice versa can be achieved
straightforward in the hybrid system. In Fig.6(d), we further give much more
bigger coupling strengths $\beta _{2}$ ($\beta _{2}=5\beta _{1}$) that
influence the absorption and dispersion of the QD. Obviously, the absorption
and dispersion are enhanced with bigger $\beta _{2}$ that induce distinct
Fano resonance and optical propagation, respectively.
\begin{figure}[tbp]
\includegraphics[width=9cm]{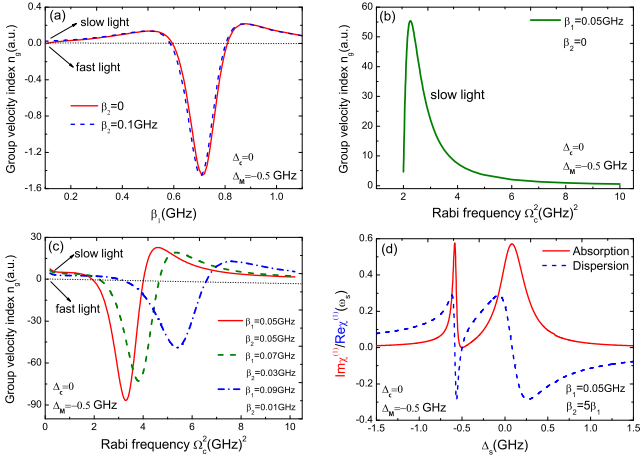}
\caption{(a) The group velocity index $n_{g}$ as a function of $\protect%
\beta _{1}$ under the condition of $\protect\beta _{2}=0$ and $\protect\beta %
_{2}=0.1$ GHz, respectively. (b) The group velocity index $n_{g}$ versus the
Rabi frequency $\Omega _{pu}^{2}$ in the case of $\protect\beta _{1}=0.05$
GHz and $\protect\beta _{2}=0$. (c) The group velocity index $n_{g}$ \textit{%
vs.} Rabi frequency $\Omega _{pu}^{2}$ under other three conditions. (d) The
absorption and dispersion of the QD in much more bigger coupling strengths $%
\protect\beta _{2}$ ($\protect\beta _{2}=5\protect\beta _{1}$). The detuning
conditions are $\Delta _{c}=0$ and $\Delta _{M}=-0.5$ GHz.}
\end{figure}

\subsection{Case C: the exciton-pump field detuning off-resonant $\Delta
_{c}\neq 0$\ under coupled Majorana modes $\protect\epsilon _{M}\neq 0$}

In this case, we first adjust the detuning $\Delta _{c}$ from $\Delta _{c}=0$
to $\Delta _{c}=0.5$ GHz under coupled Majorana modes $\epsilon _{M}\neq 0$.
Figure 7(a) gives the probe absorption spectra with several different QD-MFs
coupling strengths $\beta _{2}$ for $\beta _{1}=0.05$ GHz under the
condition of $\Delta _{M}=-0.5$ GHz and $\Delta _{c}=0.5$ GHz, which shows
more remarkable Fano resonance and the intensity is also enhanced
simultaneously. Figure 7(b) is the detail parts of the left peaks in
Fig.7(a). Compared with Fig.3(a) and Fig.5(a), we find the splitting
distance in Fig.7(a) is the sum of the splitting distance in Fig.3(a) and
Fig.5(a), i.e., the splitting distance is $\Delta _{c}-\Delta _{M}$. Figure
7(c) plots the dispersion of the probe light with four coupling strengths $%
\beta _{2}$ and Fig. 7(d) is its detail. Therefore, the remarkable fast to
slow light (or vice versa) can also be achieved in this condition. \
Moreover, when we make a comparison between the conditions in Case C and
Case A (or Case B), we find the full width at half maximum (FWHM) is
narrower in Case C than in Case A (or Case B), as well as the stronger
intensity. So, the sideband peak induced by coupled Majorana modes coincides
with sharp peaks induced by pump off-resonant $\Delta _{c}\neq 0$, which
makes the coherent interaction of QD-MF more strong.
\begin{figure}[tbp]
\includegraphics[width=9cm]{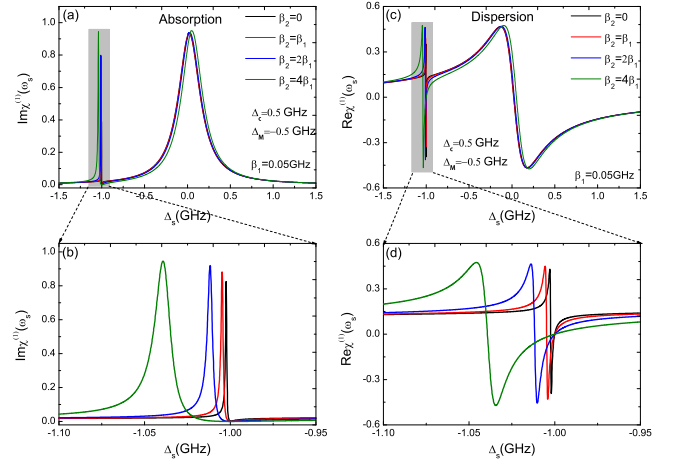}
\caption{(a) The probe absorption spectra for several different QD-MFs
coupling strengths $\protect\beta _{2}$ at $\protect\beta _{1}=0.05$ GHz,
and (b) is the detail parts of the left peaks. (c) The dispersion of the
probe light with four coupling strengths $\protect\beta _{2}$ and (d) is its
detail. The detuning conditions are $\Delta _{M}=-0.5$ GHz and $\Delta
_{c}=0.5$ GHz,}
\end{figure}

We then change the exciton-pump field detuning $\Delta _{c}$ from $\Delta
_{c}=-1.5$ GHz to $\Delta _{c}=1.5$ GHz under $\Delta _{M}=-0.5$ GHz, the
scenario of absorption becomes completely different. Figure 8 shows the
probe absorption spectra as a function of $\Delta _{s}$ with fixed pump
intensity $\Omega _{pu}^{2}=0.005($GHz$)^{2}$, which experiences the switch
from Fano resonance to EIT to Fano resonance with the change of the detuning
$\Delta _{c}$. It is obvious that these Majorana-induced resonances have a
Fano-like shape that varies with the detuning $\Delta _{c}$. When the
drive-resonance detuning $\Delta _{c}=\Delta _{M}$, the Fano-like resonance
changes into a symmetric Lorentzian-shaped absorption peak (i.e., MMIT). In
addition, we find that the transparency windows (the absorption dip
approaches zero) still locate at $\Delta _{s}=-1.0$ GHz. However, the
Lorentz peaks vary with the change of the detuning $\Delta _{c}$ and locate
at $\Delta _{c}+\Delta _{M}$. This behavior may be ascribed to the
off-resonant coupling between the QD and MFs. In addition, the probe
absorption splits into two resonances, known as the Autler-Townes (AT)
splitting, is also observed in strongly driven quantum dot system \cite{XuX}%
, where the probe absorption spectra display symmetrical splitting when the
pump is on resonance and show unsymmetric splitting at off-resonance. When
we consider a pair of MFs coupled to the QD, the evolution of the absorption
changes significantly and the probe absorption spectra are very different
from a single QD system. This further demonstrate the role of MFs in the
hybrid system, and MFs provide a quantum channel to affect the probe
absorption. Obviously, the absorption spectra can be modified effectively
via the off-resonant coupling between the QD and MFs.
\begin{figure}[tbp]
\includegraphics[width=9cm]{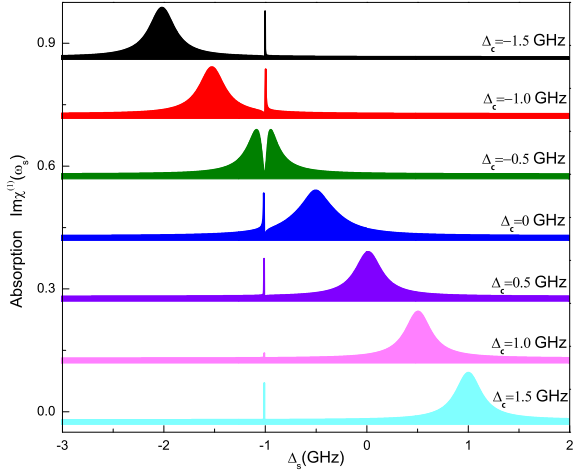}
\caption{The probe absorption spectra as a function of $\Delta _{s}$ with
fixed pump intensity $\Omega _{pu}^{2}=0.005($GHz$)^{2}$ for seven different
detuning $\Delta _{c}$ under $\Delta _{M}=-0.5$ GHz.}
\end{figure}

On the other hand, we discuss the variation of the probe absorption spectra
under two conditions of $\Delta _{c}=0.5$ GHz and $\Delta _{c}=-0.5$ GHz
with change the Majorana-pump field detuning $\Delta _{M}$ from $\Delta
_{M}=-1.5$ GHz to $\Delta _{M}=1.5$ GHz, respectively. Figure 9(a) plots a
series of probe absorption spectra under $\Delta _{c}=0.5$ GHz with several
different $\Delta _{M}$. In this condition, Fano resonance vary
significantly and Lorentz-like peaks locate at $\Delta _{s}=0$, except $%
\Delta _{M}=0.5$ GHz presenting EIT. The location of the transparency
windows (i.e., the absorption dip approaches zero) vary with the change of $%
\Delta _{M}$, and we find that the distance of the splitting is $\Delta
_{M}-\Delta _{c}$. Figure 9(b) gives a series of probe absorption spectra
under $\Delta _{c}=-0.5$ GHz with seven different $\Delta _{M}$. In this
situation, the evolution of Fano resonances are also remarkable and
Lorentz-like peaks locate at $\Delta _{s}=-1.0$ GHz, but the phenomenon of
EIT locating at $\Delta _{M}=-1.0$ GHz, and the distance of the splitting of
two peaks is also $\Delta _{M}-\Delta _{c}$. Compared with Fig. 9(a) and
Fig. 9(b), we find probe absorption spectra at $\Delta _{M}=\Phi $ and $%
\Delta _{M}=-\Phi $ ($\Phi $ indicates numerical value) presenting mirror
symmetry, such as $\Delta _{M}=1.5$ GHz and $\Delta _{M}=-1.5$ GHz, and so
on.
\begin{figure}[tbp]
\includegraphics[width=9cm]{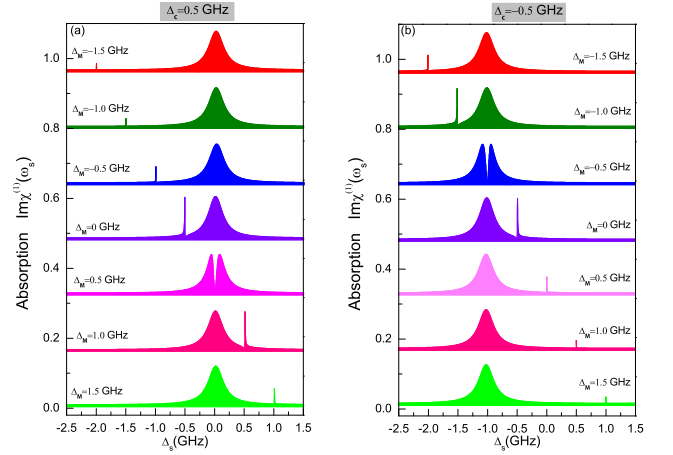}
\caption{A series of probe absorption spectra under $\Delta _{c}=0.5$ GHz
with several different $\Delta _{M}$. (b) A series of probe absorption
spectra under $\Delta _{c}=-0.5$ GHz with seven different $\Delta _{M}$.}
\end{figure}

In Fig. 10, we plot the group velocity index $n_{g}$ of probe laser under
different detuning regimes. In Fig.10(a), we investigate the group velocity
index $n_{g}$ as a function of $\beta _{1}$ for $\beta _{2}=0$ and $\beta
_{2}=0.1$ GHz under $\Delta _{c}=0.5$ GHz and $\Delta _{M}=-0.5$ GHz,
respectively, which presents the conversion from fast to slow light. When $%
\beta _{1}<0.1$ GHz, the experience\ of the group velocity index $n_{g}$ is
a little difference, while when $\beta _{1}>0.1$ GHz, the two curves are
approximate coincide. Figure 10(b) shows the group velocity index $n_{g}$ of
probe laser versus the Rabi frequency $\Omega _{pu}^{2}$ of the pump field
corresponding to Fig.8, and we only plot four curves at four different
detuning $\Delta _{c}$ under $\Delta _{M}=-0.5$ GHz. Obviously, the group
velocity index $n_{g}$ undergo advance to delay corresponding to fast to
slow light. Figure 10(c) plots the group velocity index $n_{g}$ versus the
Rabi frequency $\Omega _{pu}^{2}$ corresponding to Fig.9(a), and we give two
curves at $\Delta _{M}=-0.5$ GHz (the blue one) and $\Delta _{M}=0.5$ GHz
(the red one) under $\Delta _{c}=0.5$ GHz, which indicates the conversion
from fast to slow light and slow to fast light, respectively. Figure 10(d)
plots the group velocity index $n_{g}$ versus the Rabi frequency $\Omega
_{pu}^{2}$ corresponding to Fig.9(b), and we give two curves at $\Delta
_{M}=-0.5$ GHz (the blue one) and $\Delta _{M}=0.5$ GHz (the red one) under $%
\Delta _{c}=-0.5$ GHz. Compared with Fig. 10(c), although it can also obtain
the conversion from fast to slow light and vice versa, the process of
evolution is a little different from in Fig.10(c). Thus, with controlling
different detuning regimes, the fast-to-slow light or vice versa can be
achieved straightforward in the hybrid system.
\begin{figure}[tbp]
\includegraphics[width=9cm]{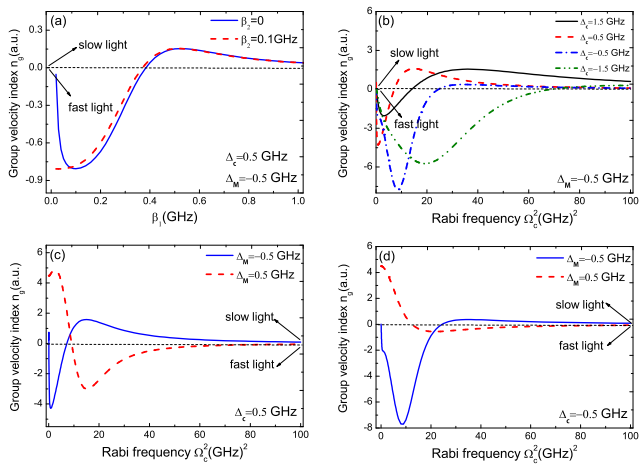}
\caption{(a) The group velocity index $n_{g}$ as a function of $\protect%
\beta _{1}$ for $\protect\beta _{2}=0$ and $\protect\beta _{2}=0.1$ GHz
under $\Delta _{c}=0.5$ GHz and $\Delta _{M}=-0.5$ GHz, respectively, (b)
The group velocity index $n_{g}$ of probe laser versus the Rabi frequency $%
\Omega _{pu}^{2}$ of the pump field for four different detuning $\Delta _{c}$
under $\Delta _{M}=-0.5$ GHz. (c) The group velocity index $n_{g}$ versus
the Rabi frequency $\Omega _{pu}^{2}$ under $\Delta _{M}=-0.5$ GHz and $%
\Delta _{M}=0.5$ GHz with $\Delta _{c}=0.5$ GHz. (d) The group velocity
index $n_{g}$ versus the Rabi frequency $\Omega _{pu}^{2}$ at $\Delta
_{M}=-0.5$ GHz and $\Delta _{M}=0.5$ GHz for under $\Delta _{c}=-0.5$ GHz. }
\end{figure}

\section{Conclusion}

We have investigated the Fano resonance and coherent optical propagation
properties in the hybrid QD-S/S ring device, which includes a QD driven by
two-tone fields coupled to a couple of MFs emerging in the ends of the
nanowire of the hybrid S/S ring device. We found Majorana modes induced
transparency, the Fano resonances and their related propagation properties
such as fast and slow light effects can achieved under different parameter
regimes, such as the exciton-pump field detuning $\Delta _{c}$, the
Majorana-pump field detuning $\Delta _{M}$, and the QD-MFs coupling
strengths $\beta _{1}$ and $\beta _{2}$. With controlling the detuning of $%
\Delta _{c}$ and $\Delta _{M}$, a series of asymmetric Fano line shapes can
appear and the positions of the Fano line shapes are closely related to the
two detuning. In addition, the fast-to-slow light or vice versa can be
achieved in the hybrid system by controlling different detuning regimes. The
scheme proposed here may provide potential applications in all-optically
controlled quantum computing based on MFs in solid-state devices.

\section*{Acknowledgement}

Hua-Jun Chen is supported by the National Natural Science Foundation of
China (Nos:11647001 and 11804004) and Anhui Provincial Natural Science
Foundation (Nos:1708085QA11).

\end{document}